\documentclass[article,showpacs,showkeys,superscriptaddress]{revtex4}

\usepackage{amsmath}
\usepackage{bbold}
\usepackage{amsfonts}
\usepackage{amssymb}
\usepackage{pbsi}
\usepackage[T1]{fontenc}
\usepackage{xcolor}
\usepackage{graphicx}
\usepackage{subfig}
\usepackage{float}
\usepackage{color}
\usepackage{hyperref}
\hypersetup{colorlinks = true,
	linkcolor = red,
	anchorcolor = blue,
	citecolor = blue,
	filecolor = blue,
	urlcolor = blue}

\begin{document}

\title{Electromagnetic plasma wave modes propagating along light-cone coordinates}

\author{Felipe A. Asenjo}
\email{felipe.asenjo@uai.cl}
\affiliation{Facultad de Ingenier\'ia y Ciencias,
Universidad Adolfo Ib\'a\~nez, Santiago 7491169, Chile.}
\author{Swadesh M. Mahajan}
\email{mahajan@mail.utexas.edu}
\affiliation{Institute for Fusion Studies, The University of Texas at Austin, Texas 78712, USA.}

\date{\today}

\begin{abstract}
We present new electromagnetic plasma wavepacket solutions that propagates in one time and one space coordinates. Differently to the usual plane wave solution, which is written in terms of separation of variables, all our solutions are along the light-cone coordinates. This allow us to find several new  solutions whose functionality properties rely on the conditions imposed on the choice for their light-cone coordinates dependence. The presented wavepacket solutions are constructed in terms of multiplications of Airy functions, Parabolic cylinder functions, Mathieu functions, or Bessel functions.
We thoroughly analyze the case of a double Airy solution, which have new electromagnetic properties, as a defined wavefront, and velocity faster than the electromagnetic plane wave counterpart solution. It is also mentioned how more general structured wavepackets can be constructed from these new solutions.
\end{abstract}

%PACS numbers

%keywords

\maketitle

\section{Introduction}

In this brief work we discuss exact  modes for electromagnetic plasma waves that can be constructed by forcing the propagation along  light-cone coordinates. Therefore, all the solutions are one-dimensional in space.
These new modes pertain to a broad family of solutions that are very different from the usual propagating plane wave, complementing other plasmas modes already found with similar properties \cite{Li,Mino,winkler,asenjo1,asenjo2}.   In particular, the solutions presented here, can also   bring new insights to  previous known  propagating modes studied in laser-plasmas interactions, such as Airy-Bessel beams \cite{chong,Ren,Ersoy,Porat,Pawar}, Bessel-Gaussian beams \cite{kant,Fallah,Hafizi,Hafizi2, Kulya} or Bessel-Bessel beams \cite{salamin,salamin2,Kotlyar}
Laguerre-Gaussian beams \cite{Wang,culfa,Baumann,kad,patil}, among others.
These new solutions, with their unique structures, are, therefore, expected to be realizable in laboratory.

  Characteristic features of all those known wavepacket solutions emerge because the wave is extended in different spatial directions. However, the currently explored  1+1 dimensional wavepackets, to the best of our knowledge,  correspond to a new set of wavepacket modes of propagation. These solutions are found by solving the simplest but general wave equation for electromagnetic plasma waves, now written in terms of light-cone coordinates. Interestingly, the resulting equation can be solved in terms of a variety of functions with complex structures (a superposition of plane waves with different weight functions). In general, there exists an asymmetry between the propagation along the two light-cone coordinates; the wavepackets are not just a simple change of variables in the plane wave solution.

We review first the relativistic theory and derivation of the electromagnetic plasma waves. We will deal with the simplest manifestation of a cold relativistic plasma with a single dynamic species (for instance, in a neutralizing ion background). Wave propagation in such a plasma with density $n$ and four-velocity $U^\mu$, is modeled by the continuity equation
$\partial_\mu(n U^\mu)=0$ (where $\partial_\mu$ is the partial covariant derivative), 
the momentum equation 
\begin{equation}
    m U_\nu\partial^\nu  U^\mu=q F^{\mu\nu}U_\nu\, ,
    \label{momentum}
\end{equation}
and the Maxwell equations
\begin{equation}
    \partial_\nu F^{\mu\nu}=q n U^\mu\, ,
    \label{maxwell}
\end{equation}
where $F^{\mu\nu}=\partial^\mu A^\nu-\partial^\nu A^\mu$ is the Faraday tensor with $A^\mu$ as the electromagnetic four-potential. Here, $m$ ($q$) is the mass (charge) of the plasma fluid constituent, and the velocity of light $c=1$.  

The simplest solution of the above system happens to be for electromagnetic plasma waves. When $A^\mu=-(m/q)U^\mu$, the momentum equation \eqref{momentum} is identically solved for an homentropic plasma \cite{asenjosusy}. Then, the 
continuity equation reduces to the Lorenz gauge $\partial_\mu A^\mu=0$ (for a cold constant density plasma). The resulting Maxwell equation \eqref{maxwell} 
\begin{equation}
    \partial_\nu\partial^\nu A^\mu+\omega_p^2 A^\mu=0\, ,
    \label{waveeq}
\end{equation}
where $\omega_p=\sqrt{q^2 n/m}$ is the (constant) plasma frequency, is the starting point for this enquiry.   For transverse electromagnetic polarization (propagation along $x$, the electromagnetic fields in the $y$-$z$ plane) for instance, we will need to solve for $A^z(t,x)$, obeying 
\begin{equation}
    \left(\frac{\partial^2}{\partial t^2}-\frac{\partial^2}{\partial x^2}\right)A^z+\omega_p^2 A^z=0\, .
    \label{waveeq2}
\end{equation}
whereas all the solutions satisfy the Lorenz gauge by construction.

It is customary to exactly solve Eq.~\eqref{waveeq2} in terms of transversal plane waves, $A^z(t,x)=\exp(i\omega t-i kx)$, with the dispersion relation for electromagnetic waves $\omega^2=k^2+\omega_p^2$. This dispersion relation also establishes a specific sub-luminal propagation velocity of the plasma wave. Its  group velocity, $v_g=\partial\omega/\partial k=(1+\omega_p^2/k^2)^{-1/2}<1$, is completely determined by the plasma content of the system, through $\omega_p$.
This above basic solution is conventionally obtained via separation of variables.

However, we plan to explore here another (possible unique) class of solutions in terms  
of the light-cone coordinate variables $\eta=x-t$ and $\xi=x+t$. This paper is structured as follow. In Sec.~\ref{sec1} we show how these new solution can be obtained from Eq.~\eqref{waveeq2}. In Sec.~\ref{sec2} we discuss the velocity of propagation of such waves. In Sec.~\ref{sec3} we study the specific case of the double Airy solution, and to finally in Sec.~\ref{sec4} we discuss our findings.

\section{Wave equation in light-cone coordinates and their solutions}
\label{sec1}

We will, now, seek separable solutions of the form $A^z(t,x)=A^z(\rho,\chi)=f(\rho)g(\chi)$, in terms of variables  $\rho=\rho(\eta,\xi)$ and $\chi=\chi(\eta,\xi)$ that are some combinations of the light cone coordinates $\eta=x-t$ and $\xi=x+t$. The solutions are separable in terms of a functionality form, and not in terms of variables.
We will find that \eqref{waveeq2} will dictate the choices for $\rho=\rho(\eta,\xi)$ and $\chi=\chi(\eta,\xi)$ that will allow analytically expressible solutions.

%Let us consider a solution in terms of an arbitrary separation of variables $A^z(t,x)=A^z(\rho,\chi)=f(\rho)g(\chi)$, where for now $f$ and $g$ are arbitrary functions of $\rho=\rho(\eta,\xi)$ and $\chi=\chi(\eta,\xi)$, and where  $\eta=x-t$ and $\xi=x+t$ are the light-cone coordinates of the system.
%

After variable change, the wave equation \eqref{waveeq2} becomes
\begin{eqnarray}
\left[\partial_\eta\rho\partial_\xi\rho\frac{\partial^2}{\partial\rho^2}+\left(\partial_\eta\rho\partial_\xi\chi+\partial_\xi\rho\partial_\eta\chi\right)\frac{\partial^2}{\partial\chi\partial\rho}+\partial_\eta\chi\partial_\xi\chi\frac{\partial^2}{\partial\chi^2}\right]fg-\frac{\omega_p^2}{4}fg=0\, .
\label{waveeq3}
\end{eqnarray}
%As we are interested in a solution with separation 
%of variables $(\rho, \chi)$,  

The choice 
\begin{eqnarray}
    \rho(\eta,\xi)&=&\theta(\eta)+\phi(\xi)\, ,\nonumber\\
    \chi(\eta,\xi)&=&\theta(\eta)-\phi(\xi)\, ,
\end{eqnarray}
will convert  Eq.~\eqref{waveeq3} to an interesting form
\begin{equation}   \partial_\eta\theta\partial_\xi\phi\left(\frac{\partial_\rho^2 f}{f}-\frac{\partial_\chi^2 g}{g}\right) -\frac{\omega_p^2}{4}=0\, ,
   \label{waveeq4}
\end{equation}
that represents a more explicit formulation for studying wave propagation along the light-cone coordinates (the original wave equation is, of course, \eqref{waveeq2}).
Eq.~\eqref{waveeq4} will be our fundamental basis for exploring the class of solutions that we promised. The trick is to choose $\theta(\eta), \phi(\xi)$ so that \eqref{waveeq4} splits into two analytically solvable equations for $f$ and $g$. Of course all solutions of \eqref{waveeq4} are solutions of  \eqref{waveeq2}.
%In the following, several distinct solutions (very different to plane waves) are shown to satisfy Eq.~\eqref{waveeq4} [and thus, Eq.~\eqref{waveeq2}].
%
We will now go systematically making appropriate choices to find explicit solutions. The first choice leads to what we call a Double Airy mode.

\subsection{Double Airy mode}
\label{douairy}

The simplest and non-trivial \cite{note}  solution of Eq.~\eqref{waveeq4} is achieved by assuming that $f$ and $g$ are Airy functions, ${\partial_\rho^2 f}=\rho {f}$, and ${\partial_\chi^2 g}=\chi {g}$. In this case, ${\partial_\rho^2 f}/{f}-{\partial_\chi^2 g}/{g}=\rho-\chi=2\phi$, and then Eq.~\eqref{waveeq4} becomes
\begin{equation}
   \partial_\eta\theta\, \partial_\xi\phi^2-\frac{\omega_p^2}{4}=0\, ,
   \label{waveeq6}
\end{equation}
which is solved by  $\theta=\eta/\alpha^2$, and $\phi=\alpha\, \omega_p\sqrt{\xi}/2$, with the arbitrary parameter $\alpha$. Explicitly, the solution of Eq.~\eqref{waveeq2}, in terms of space and time coordinates, becomes
\begin{eqnarray}
A^z(t,x,\omega_p,\alpha)&=&\mbox{Ai}\left(\frac{x-t}{\alpha^2}+\frac{\alpha\omega_p}{2}\sqrt{x+t} \right)\mbox{Ai}\left(\frac{x-t}{\alpha^2}-\frac{\alpha\omega_p}{2}\sqrt{x+t} \right)\, .
\label{Azdoubleairy}
\end{eqnarray}
This obviously complex solution has been previously obtained in the context of 
waveguides \cite{otk}. Interestingly, this double Airy electromagnetic plasma wavepacket presents an  asymmetry between the two light-cone coordinates. First, by the different functionality forms between the two of them, and secondly, by their different ponderation through the arbitrary $\alpha$ constant.  
For instance, as $\alpha$ decreases, the wavepacket propagation tends to align with the light-cone coordinate $\eta=x-t$. Thus, this wavepacket can propagate arbitrarily close to the speed of light, independent of the value of the plasma frequency $\omega_p$. This behavior is very different to the usual electromagnetic plasma plane wave (with its specific dispersion relation), and it will be discussed in Sec.~\ref{sec3}.

The solution could, indeed, be also obtained in terms of any combinations of Airy functions of the first and second kind. Both $\mbox{Bi}$, or a combinations of $\mbox{Ai}$ and $\mbox{Bi}$ will do.

\subsection{Double Parabolic Cylinder mode}

A different solution of Eq.~\eqref{waveeq4} can be obtained if both $f$ and $g$ were to satisfy the same generalized Weber equation $W''(y)=(a y^2+b y-n)W(y)$ (where $a$, $b$, and $n$ are constants) which is solved in terms of parabolic cylinder functions. In this case, Eq.~\eqref{waveeq4} will be reduced to
\begin{equation}
\partial_\eta\left(a\theta^2+b\theta\right)\, \partial_\xi\phi^2-\frac{\omega_p^2}{4}=0\, .
   \label{waveeq7}
\end{equation}
This equation is solved by 
\begin{equation}
 \theta=-b/(2a)+\sqrt{b^2/(4a^2)+\eta/(a\alpha^2)}, \quad  \phi=\alpha\, \omega_p\sqrt{\xi}/2,
 \end{equation}
 where $\alpha$ is an arbitrary number. Collecting the pierces together, the complete solution is 
 \begin{eqnarray}
A^z(t,x,\omega_p,\alpha,a,b,n)&=&D_{p}\left(-\frac{b}{2a}+\sqrt{\frac{b^2}{4a^2}+\frac{x-t}{a\alpha^2}}+\frac{\alpha\omega_p}{2}\sqrt{x+t} \right)\nonumber\\
&&\times D_{p}\left(-\frac{b}{2a}+\sqrt{\frac{b^2}{4a^2}+\frac{x-t}{a\alpha^2}}-\frac{\alpha\omega_p}{2}\sqrt{x+t} \right)\, ,
\end{eqnarray}
where $p=n/(2\sqrt{a})+b^2/(8a^{3/2})-1/2$ is the index of the parabolic cylinder function $D_p$. Anew, in this solution (and in the following ones) there is an asymmetry between the two light-cone coordinates, depending on arbitrary parameters.

% The solutions of this equation can be written in terms of
%parabolic cylinder function $D_{p}(\sqrt{2}a^{1/4}y+b/\sqrt{2 a^{3/2}})$, with $p=n/(2\sqrt{a})+b^2/(8a^{3/2})-1/2$.
%
%For this case, ${\partial_\rho^2 f}/{f}-{\partial_\chi^2 g}/{g}=2\phi(2a\theta+b)$, and therefore, Eq.~\eqref{waveeq4} becomes
%\begin{equation}
%   \partial_\eta\left(a\theta^2+b\theta\right)\, \partial_\xi\phi^2-\frac{\omega_p^2}{4}=0\, .
%   \label{waveeq7}
%\end{equation}
%The solution of this equation is in terms of $\theta=-b/(2a)+\sqrt{b^2/(4a^2)+\eta/(a\alpha^2)}$, and $\phi=\alpha\, \omega_p\sqrt{\xi}/2$, where $\alpha$ is arbitrary.
%In this case, the solution of Eq.~\eqref{waveeq2}, written in terms of space and time coordinates, is
%\begin{eqnarray}
%A^z(t,x,\omega_p,\alpha,a,b,n)&=&D_{p}\left(-\frac{b}{2a}+\sqrt{\frac{b^2}{4a^2}+\frac{x-t}{a\alpha^2}}+\frac{\alpha\omega_p}{2}\sqrt{x+t} \right)\nonumber\\
%&&\times D_{p}\left(-\frac{b}{2a}+\sqrt{\frac{b^2}{4a^2}+\frac{x-t}{a\alpha^2}}-\frac{\alpha\omega_p}{2}\sqrt{x+t} \right)\, ,
%\end{eqnarray}
%where now the electromagnetic plasma wavepacket depends also on the kind of the parabolic cylinder solution, through $p$. 
%

\subsection{Double Mathieu mode}

Continuing this journey, we will now show that if both  $f$ and $g$ satisfy the Mathieu equation $M''(y)+(a-b\cos(y))M(y)=0$ (with solutions $M_{a,b}(y)$; $a$ and $b$ are constants) \cite{bender},  
Eq.~\eqref{waveeq4} reduces to
 \begin{equation}
   \partial_\eta(\cos\theta) \,\partial_\xi(\cos\phi)-\frac{\omega_p^2}{8b}=0\, ,
   \label{waveeqMM}
\end{equation}
which, in turn demands
 \begin{equation}
 \cos\theta=\alpha\eta, \quad \cos\phi=\omega_p^2 \xi/(8b\alpha),
\end{equation}
 for an arbitrary $\alpha$. Consequently
\begin{eqnarray}    A^z(t,x,\omega_p,\alpha,a,b)&=&M_{a,b}\left(\arccos(\alpha(x-t))+\arccos\left(\frac{\omega_p^2}{8b\alpha}(x+t)\right) \right)\nonumber\\
&&\times M_{a,b}\left(\arccos(\alpha(x-t))-\arccos\left(\frac{\omega_p^2}{8b\alpha}(x+t)\right) \right)\, .
\end{eqnarray}

\subsection{Double Modified Mathieu mode}

On the other hand, if  $f$ and $g$ were to satisfy the modified Mathieu equation, $M_M''(y)-(a-b\cosh(y))M_M(y)=0$ (with solutions $M_{M,a,b}(y)$), the transformed Eq.~\eqref{waveeq4} 
becomes
% ${\partial_\rho^2 f}/{f}-{\partial_\chi^2 g}/{g}=a-b\cosh\rho-(a-b\cosh\chi)=-2b\sinh\theta\sinh\phi$. 

\begin{equation}
   \partial_\eta(\cosh\theta) \,\partial_\xi(\cosh\phi)-\frac{\omega_p^2}{8b}=0\, ,
   \label{waveeqMMM}
\end{equation}
which is satisfied for $\cosh\theta=\alpha\eta$, and $\cosh\phi=\omega_p^2 \xi/(8b\alpha)$, for an arbitrary $\alpha$. 
The explicit solution, then, is 
\begin{eqnarray}    A^z(t,x,\omega_p,\alpha,a,b)&=&M_{M,a,b}\left(\mbox{arccosh}(\alpha(x-t))+\mbox{arccosh}\left(\frac{\omega_p^2}{8b\alpha}(x+t)\right) \right)\nonumber\\
&&\times M_{M,a,b}\left(\mbox{arccosh}(\alpha(x-t))-\mbox{arccosh}\left(\frac{\omega_p^2}{8b\alpha}(x+t)\right) \right)\, .
\end{eqnarray}

\subsection{Double Modified Bessel mode}

We will now work out the final example involving modified Bessel functions. Let
$\partial_\rho^2f=a\, e^{b\rho} f$, and $\partial_\chi^2g=a\, e^{b\chi} g$ (where $a$ and $b$ are constants). This special case is engineered to solutions of the form $f(\rho)=K_0(2\sqrt{a} e^{b\rho/2}/b)$ (and similar for $g$), where $K_0$ is the modified Bessel function of order 0. The solubility condition for Eq.~\eqref{waveeq4}, then, becomes 

%Therefore, $\partial_\rho^2f/f-\partial_\chi^2g/g=2a e^{b\theta}\sinh(b\phi)$, and Eq.~\eqref{waveeq4} 

\begin{equation}
   \partial_\eta(e^{b\theta}) \,\partial_\xi(\cosh(b\phi))-\frac{\omega_p^2b^2}{8a}=0\, .
   \label{waveeqKK}
\end{equation}
yielding $e^{b\theta}=\alpha\eta$, and $\cosh(b\phi)=\omega_p^2 b^2\xi/(8a\alpha)$, for arbitrary $\alpha$. The explicit spacetime solution of the original wave equation 
follows
\begin{eqnarray}    A^z(t,x,\omega_p,\alpha,a,b)&=&K_0\left(\frac{2\sqrt{a}}{b}\sqrt{\frac{\omega_p^2 b^2}{8a}(x^2-t^2)+\alpha(x-t)\sqrt{\frac{\omega_p^4 b^4}{64 a^2\alpha^2}(x+t)^2-1}} \right)\nonumber\\
&&\times K_0\left(\frac{2\sqrt{a}}{b} \sqrt{\frac{\alpha(x-t)}{ \frac{\omega_p^2 b^2}{8a\alpha}(x+t)+\sqrt{\frac{\omega_p^4 b^4}{64 a^2\alpha^2}(x+t)^2-1}}}\right)\, .
\end{eqnarray}

%\begin{eqnarray}    A^z(t,x,\omega_p,\alpha,a,b)&=&K_0\left(\frac{2\sqrt{a}}{b}\exp\left[\frac{1}{2}\ln(\alpha(x-t))+\frac{1}{2}\mbox{arccosh}\left(\frac{\omega_p^2 b^2}{8a\alpha}(x+t) \right)\right] \right)\nonumber\\
%&&\times K_0\left(\frac{2\sqrt{a}}{b}\exp\left[\frac{1}{2}\ln(\alpha(x-t))-\frac{1}{2}\mbox{arccosh}\left(\frac{\omega_p^2 b^2}{8a\alpha}(x+t) \right)\right] \right)\, ,
%\end{eqnarray}

\subsection{General wavepackets}

Notice that we have concentrated on building solutions Eq.~\eqref{waveeq4} that are explicitly expressible in terms of known special functions. Though representing highly complex structures, these   are, nonetheless, the simplest non-trivial solvers of Eq.~\eqref{waveeq4}. They all depend on an  arbitrary constant $\alpha$ which represents some arbitrariness on the scale length of the solution. In particular it is a measure of what light-cone coordinate will dominate the overall structure.

Not knowing {\it a priori} what $\alpha$ to pick, the general solution of Eq.~\eqref{waveeq4} can be constructed by the superposition of $A^z(\alpha)$ on different distributions of $\alpha$. This superposition is allowed because $A^z(\alpha)$ is a solution of the linear wave equation. Mathematically, we construct
\begin{eqnarray}
    {\mathfrak A}(t,x,\omega_p,\vec\kappa)= \int \zeta(\alpha)\, A^z(t,x,\omega_p,\alpha,\vec\kappa)\, d\alpha\, 
    \label{wavepacketG}
\end{eqnarray}
where $\vec\kappa$ represent the  set of parameter of the different previous solutions, and $\zeta(\alpha)$ is any possible continuous function depending only on $\alpha$.  

An example of this $\alpha$ averaged wavepacket will be presented for the Double Airy solution in Sec.~\ref{sec3}.

\section{Energy velocity of the solutions}
\label{sec2}

In general, the wavepackets constructed in this paper are non-monochromatic. The concept of phase or group velocity (derived from a dispersion relation for plane waves), therefore, is not strictly applicable.
The propagation velocity of the whole structure must be defined through the concept of the velocity of transport of energy of the wavepacket. This energy velocity is defined as $v_\epsilon=|\vec S|/\epsilon$, where $\vec S$ is the Poynting vector, and $\epsilon$ is the energy density of the electromagnetic field.

For the electromagnetic field characterized by $A^z(t,x)$, following relations hold:  $E(t,x)=-\partial_t A^z(t,x)$, $B(t,x)=-\partial_x A^z(t,x)$, the Poynting $|\vec S|= E B$, and the energy density is $\epsilon=E^2/2+B^2/2$.

Using the general form of $A^z(t,x)= f(\rho)g(\chi)$, we readily calculate  that the energy velocity of our solutions is 
 \begin{equation}
 v_\epsilon(t,x,\alpha)=\frac{1-\left({\partial_\xi\phi}/{\partial_\eta\theta}\right)^2\Psi}{1+\left({\partial_\xi\phi}/{\partial_\eta\theta}\right)^2\Psi}\, , 
 \label{energyvelocity}
 \end{equation}
 where
\begin{eqnarray}
    \Psi(t,x,\alpha)=\left(\frac{ g\partial_\rho f-f\partial_\chi g}{ g\partial_\rho f+f\partial_\chi g}\right)^2\, .
\end{eqnarray}
Therefore, the energy velocity is always subluminal, $v_\epsilon<1$. However, it varies in space and time, depending on how the solution $f$ and $g$ propagates along the light-cone coordinates. Whenever the term $\left({\partial_\xi\phi}/{\partial_\eta\theta}\right)^2\Psi$ approaches 0,  the energy velocity approaches the speed of light. This can be independent of the plasma frequency of the medium, and it can occur in an arbitrary fashion. In the following section we give an example of this for the Double Airy solution.

\section{Double Airy wave propagation in  detail}
\label{sec3}

In order to show some of the interesting features of these solutions propagating along the light-cone 
coordinates, we consider the simplest solution for the Double Airy wavepacket described in Sec.~\ref{douairy}.

\begin{figure}
    \centering
    \includegraphics[width=0.9\linewidth]{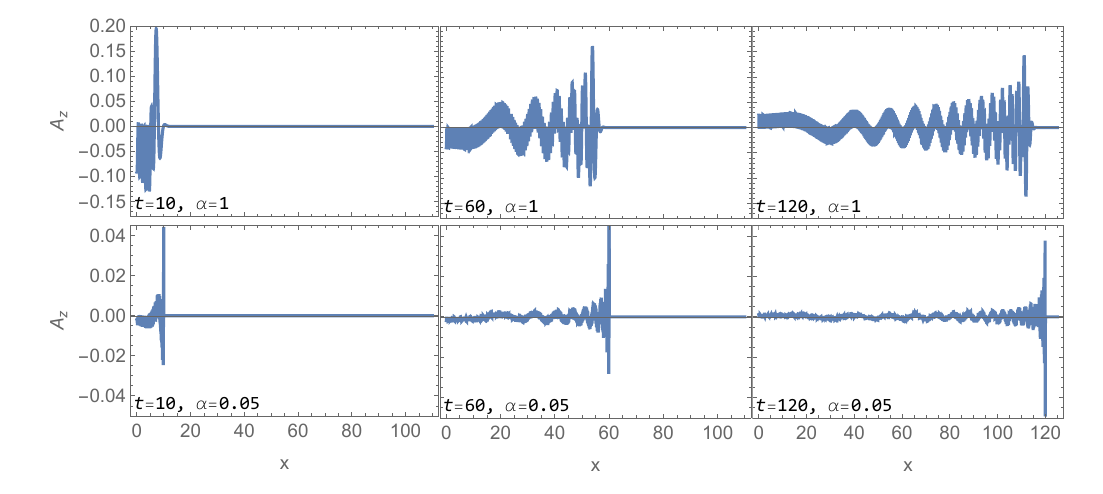}
    \caption{Double Airy solution described in Sec.~\ref{douairy} evolving for three times. Upper row for $\alpha=1$, and lower row for $\alpha=0.05$. Without loss of generality, we have chosen $\omega_p=1$.}
    \label{fig1}
\end{figure}

In Fig.~\ref{fig1}, we show the temporal evolution of this wavepacket \eqref{Azdoubleairy}  for two values of $\alpha$. We can see how this wavepacket propagates as a strong wavefront, that later decays with an exponential envelope (because of its Airy nature). We can see how the wavepacket becomes more localized as $\alpha$ gets smaller.

\begin{figure}
    \centering
    \includegraphics[width=0.7\linewidth]{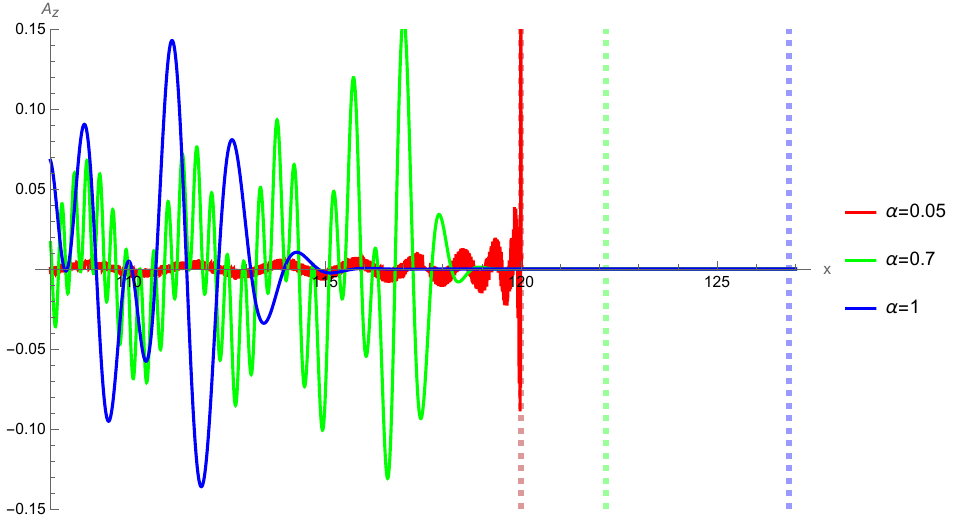}
    \caption{Double Airy solution described in Sec.~\ref{douairy} for $t=120$ for three $\alpha$ values (and $\omega_p=1$). Vertical lines represent their respective wavefront positions.}
    \label{fig2}
\end{figure}

The way that the wavefront of the wavepacket \eqref{Azdoubleairy}
moves forward depends on $\alpha$. The ``location of" this wavefront is given by the maximum maximorum of $\partial_xA_z=0$, and it evolves in time.
The position of the wavefront is thus given by 
\begin{equation}
    x_{\mbox{\tiny{WF}}}(t,\alpha)=t+\frac{\alpha^6\omega_p^2}{8}+\alpha^2 {\beta}+\sqrt{\frac{\alpha^6\omega_p^2}{2}t+\frac{\alpha^{12}\omega_p^4}{64}+\frac{\alpha^8 \beta\omega_p^2}{4}}\, ,
    \label{positionwavefront}
\end{equation}
where $\beta\approx-1.01879$ is the first zero of the derivative of the Airy function. Therefore, its wavefront velocity is superluminal
\begin{equation}
    v_{\mbox{\tiny{WF}}}(t,\alpha)=\frac{d x_{\mbox{\tiny{WF}}}}{dt}=1+\frac{\alpha^6\omega_p^2}{4}\left(\frac{\alpha^6\omega_p^2}{2}t+\frac{\alpha^{12}\omega_p^4}{64}+\frac{\alpha^8 \beta\omega_p^2}{4}\right)^{-1/2}\, ,
    \label{wavefrontvelocity}
\end{equation}
in the same sense that the phase velocity for simple plasma plane wave solution is superluminal.  As $\alpha^2$ gets smaller and the time evolves, the  wavepacket becomes more localized and its wavefront velocity approaches  the speed of light $v_{\mbox{\tiny{WF}}}(t,\alpha)\approx 1+\alpha^3\omega_p/(\sqrt{8t})$. 
This is exemplified in Fig.~\ref{fig2} where the Double Airy solution is shown for $t=120$, for three different $\alpha$ values. The vertical lines represent the position of the wavefront \eqref{positionwavefront}. This wavefront velocity reduces as $\alpha$ decreases, effectively describing the position of the shock front for smaller $\alpha$ values (and the choice $\omega_p=1$).
In this way, as $\alpha$ gets smaller, this wavepacket approaches its behavior to a light shock wave. 

Similarly, we can evaluate the energy velocity \eqref{energyvelocity} of this wavepacket. 
First, notice that for this solution,
$\left({\partial_\xi\phi}/{\partial_\eta\theta}\right)^2=\alpha^6\omega_p^2/(4\xi)$. Therefore, for a small $\alpha$ value, the subluminal energy velocity approaches the speed of light in the form
\begin{equation}
    v_\epsilon(t,x,\alpha)\approx {1-\frac{\alpha^6\omega_p^2\Psi^2}{2\xi}}\, .
    \label{energyvelocityairyariry}
\end{equation}
The energy velocity of this wavepacket approaches to the speed of light faster for smaller $\alpha$, as compared to the usual plane wave solution for electromagnetic plasma waves.
 Therefore, while the transport of energy of this wavepacket is always subluminal  (remaining causal), the superluminality features are a property of the phase of the wavepacket, or its wavefront.

Notice that the wavepacket cannot move at constant speed; it belongs to the category of accelerating plasma modes. Furthermore, the wave front velocity \eqref{wavefrontvelocity} eventually can be larger than the simple phase velocity of a plane wave solution $v_f=(1+\omega_p^2/k^2)^{1/2}$. Similarly, the energy velocity \eqref{energyvelocityairyariry} can also be larger (for larger times) than the
group velocity of a plane wave solution 
$v_g=(1+\omega_p^2/k^2)^{-1/2}$.

\begin{figure}
    \centering
    \includegraphics[width=0.5\linewidth]{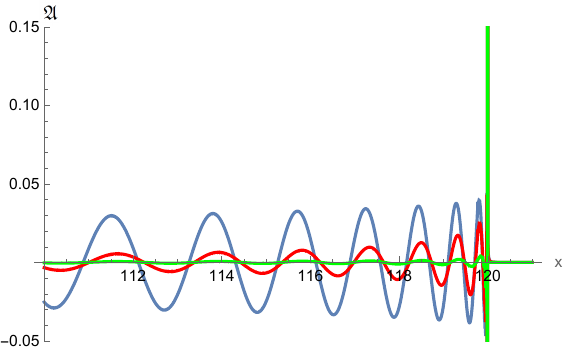}
    \caption{Wavepacket \eqref{wavepacketG} for the Double Airy solution, with $\zeta(\alpha)=\exp(-\delta\, \alpha^2)$. We consider $t=120$ and $\omega_p=1$. Blue line is for $\delta=1$. Red line is for $\delta=10$. Green line is for $\delta=100$.}
    \label{fig3}
\end{figure}

Finally, we can construct an  $\alpha$ averaged generalized wavepacket ${\mathfrak A}(t,x,\omega_p)$ of the kind represented by Eq.~\eqref{wavepacketG}. For simplicity, we can choose $\zeta(\alpha)=\exp(-\delta\, \alpha^2)$ (for some positive constant $\delta$) implying that we are focusing only on those $\alpha$ values close to 0. We are interested in such solution because of the shock wavefront properties when $\alpha$ is small. 
In Fig.~\ref{fig3} we display the constructed ${\mathfrak A}(t,x,\omega_p)$ for three values of $\delta$, for time $t=120$. For larger values of $\delta$, the shock wave behavior of  ${\mathfrak A}$ is enhanced; the intensity is concentrated more and more in the shockfront.

\section{Remarks- Perspective}
\label{sec4}

The special exact solutions ($A_z$  and ${\mathfrak A}$) for the electromagnetic fields, constructed in this paper, display rather extraordinary features associated with the propagation of electromagnetic plasma waves. Most of these predict modes where the electromagnetic energy propagates in a non-homogeneous way in the wavepackets with even well defined wavefronts.
Different from other known spatially extended wavepacket solutions, all electromagnetic fields of Sec.~\ref{sec1} do not require an extended transversal structure for their non-constant  propagation. In all cases, wavepackets do not have constant energy velocities (or any other velocity). It would seem, then, that the structured behavior of light in a plasma could be viewed as a property of the kind of coordinates in which the wavepacket propagation is studied, and not necessarily of  the number of dimensions in which the wave is studied. It is worth while to point that  $A^z$  and $\mathfrak{A}$ are exact solutions of Maxwell equations, and not approximated solutions in the eikonal limit (as most of the structured accelerating light solutions are).  All these features implies that these electromagnetic structures can exist at any energy scale of the electromagnetic spectrum.

Different solutions  exhibit asymmetry between the two light-cone coordinates. Measured by the parameter $\alpha$ (and the distribution $\zeta(\alpha)$), it labels the initial form of the wavepacket $ {\mathfrak A}(t=0,x,\omega_p,\vec\kappa)$. Consequently,  $\alpha$ is the crucial parameter that will help defining the initial condition for the  experimental realization of the wavepacket. Currently there are methods that design lasers that emit arbitrary beam profiles \cite{porat,sguo}, and thus, wavepackets as the Double Airy solution, are feasible to be constructed in laboratory. For instance, initial conditions of the solutions can be devised by the use of relativistic plasma mirrors, where the plasma dynamics is described in terms of light-cone coordinates \cite{koga,lamac}.

It is of some interest to note that the above new solutions can be generalized to three-dimensional wavepackets. By starting from wave equation \eqref{waveeq}, we can assume a solution in the form $A^z(t,x,y,z)=a^z(t,x) J_0\left(\lambda\sqrt{y^2+z^2}\right)$, where $J_0$ is the Bessel function of zeroth order, and $\lambda$ is an arbitrary constant. Substituting in  \eqref{waveeq}, then, leads to  the equation for $a^z$,
\begin{equation}
    \left(\frac{\partial^2}{\partial t^2}-\frac{\partial^2}{\partial x^2}\right)a^z+\left(\omega_p^2+\lambda^2\right) a^z=0\, ,
    \end{equation}
which is the same as  \eqref{waveeq2} but with  an effective plasma frequency, $\sqrt{\omega_p^2+\lambda^2}$. Thus, new three-dimensional wavepackets can be constructed by averaging over a distribution of $\lambda$.

On the other hand,   the same treatment of Sec.~\ref{sec3} can be performed for the other solutions of
Sec.~\ref{sec1}. For the Double Airy solution, in particular, we saw that a well defined wavefront is one of its key features. Depending on the value of $\alpha$, or the form of the averaging distribution $\zeta(\alpha)$, the waveform of the other solutions could also develop a shock front.

Finally, one must speculate whether the exotic  electromagnetic structures explored in this study are  potentially realizable in the laboratory. It seems that it may be so because  several other plasma wavepackets have been observed in laser-plasma experiments.   However,  the current linear analysis for the wavepackets excludes the nonlinear dominant effects present in the structured beams propagation in high-intensity laser-plasma interactions. This could (likely would) alter the structure and stability of the derived wavepackets. But, due to the asymmetry of the two light-cone coordinates of the presented linear solutions, it is possible that any nonlinear solution also will retain such feature, allowing us to be able to find generalizations of nonlinear solutions where space and time are treated in an asymmetrical form.

Nevertheless, these solutions define a path to find their nonlinear counterpart after experimenting with somewhat low intensity beams. Another point to keep in mind is that for circularly polarized beams, the main manifestation of the nonlinear relativistic effects is to change $\omega_p^2$ to $\omega_p^2/\gamma$; this could result in cosmetic readjustments leaving the structure mostly intact.
 
In the same spirit, all the presented wave packet solutions can be used as new test problems for PIC code simulations, as is usually do with other plasma waves \cite{kilian,Raynaud,Markidis}. We plan to handle some of these problems in near future.

\begin{acknowledgements}
FAA thanks to FONDECYT grant No. 1230094 that partially supported this work. SMM's work was performed under the US Department of Energy Award\# DE-FG02-04ER54742.

 \end{acknowledgements}


\begin{thebibliography}{28}
\bibitem{note} The trivial solution is plane waves. This is obtained for $f$ and $g$ being sinusoidal functions.
\end{thebibliography}
\end{document}